\documentclass[structabstract]{aa}  
\usepackage{graphicx}
\usepackage{txfonts}
\usepackage{natbib}
\bibpunct[; ]{(}{)}{;}{a}{}{,}
\begin{document}
\title{A ground-based $K_S$-band detection of the thermal emission 
from the transiting exoplanet WASP-4b\thanks{Based on observations 
collected at the European Southern Observatory, Chile. Programme 
083.C-0528.}}

\titlerunning{$K_S$-band thermal emission from WASP-4b.}

\author{
C. C\'{a}ceres\inst{1,2}
\and
V. D. Ivanov\inst{2}
\and
D. Minniti\inst{1}
\and
A. Burrows\inst{3}
\and
F. Selman\inst{2}
\and
C. Melo\inst{2}
\and
D. Naef\inst{4}
\and
E. Mason\inst{5}
\and
G. Pietrzynski\inst{6}
}

\institute{
Department of Astronomy and Astrophysics, P. Universidad Cat\'{o}lica de 
Chile. Av. Vicu\~{n}a Mackenna 4860, 7820436 Macul, Santiago, Chile
\mail{cccacere@astro.puc.cl}
\and
European Southern Observatory, Av. Alonso de C\'{o}rdova 3107, 
Casilla 19, Santiago 19001, Chile
\and
Department of Astrophysical Sciences, Princeton University,
Princeton, NJ 08544, USA
\and
Observatoire de Gen\`{e}ve, Universit\'{e} de Gen\`{e}ve, 51
Ch. des Maillettes, 1290 Sauverny, Switzerland
\and
ESA-STScI, 3700 San Martin Drive, Batlimore, MD 21218, U.S.A.
\and
Department of Astronomy, Universidad de Concepci\'{o}n, Casilla 
160-C, Concepci\'{o}n, Chile\\
}

\date{}


\abstract{Secondary eclipses are a powerful tool to measure directly
  the thermal emission from extrasolar planets, and to constrain their
  type and physical parameters.}%
{We started a project to obtain reliable broad-band measurements of
  the thermal emission of transiting exoplanets.}%
{Ground-based high-cadence near-infrared relative photometry was used
  to obtain sub-millimagnitude precision light curve of a secondary
  eclipse of WASP-4b -- a 1.12\,M$_{\rm J}$ hot Jupiter on a 1.34\,day
  orbit around G7V star.}%
{The data show a clear $\geq$10$\sigma$ detection of the planet's
  thermal emission at 2.2\,$\mu$m. The calculated thermal emission
  corresponds to a fractional eclipse depth of
  $0.185^{+0.014}_{-0.013}$\%, with a related brightness
  temperature in $K_S$ of $T_B = 1995 \pm 40$\,K, centered at $T_C =
  2455102.61162^{+0.00071}_{-0.00077}$\,HJD. We could set a limit on
  the eccentricity of $e\cos\omega=0.0027 \pm 0.0018$, compatible with
  a near-circular orbit.}%
{The calculated brightness temperature, as well as the specific models
  suggest a highly inefficient redistribution of heat from the
  day-side to the night-side of the planet, and a consequent emission
  mainly from the day-side. The high-cadence ground-based technique is
  capable of detecting the faint signal of the secondary eclipse of
  extrasolar planets, making it a valuable complement to space-based
  mid-IR observations.}

\keywords{Planetary systems - Eclipses - Stars: individual: WASP-4 - Techniques: photometric} 

\maketitle
%

\section{Introduction}
Photometry of extrasolar planetary transits allows to measure the
planetary radii and density, and to infer the nature of the planet -
gaseous or rocky. The detection of the thermal emission during the
secondary eclipses when the planet passes behind the star gives access
to additional physical parameters like the brightness temperature, an
important quantity for the comparison with more sophisticated
planetary atmospheric models with chemistry, and dynamics
\citep[i.e.][]{burrows_etal2008}. While a few secondary eclipses have
been secured from space \citep[e.g.][]{charbonneau_etal2005,
  deming_etal2005}, the detection of secondary eclipses has been very
challenging from the ground, with only a few recent secure detections
\citep[see][and references therein]{sing_lopez-morales2009,
  rogers_etal2009, croll_etal2010a, croll_etal2010b}.  We obtained
sub-millimagnitude precision ground-based relative photometry, based on
ultra-fast high-cadence near-infrared (NIR) observations, which can
detect transits down to a few millimagnitudes \citep[][hereafter
Paper I]{caceres_etal2009}. In this paper we report the detection of a
secondary eclipse of WASP-4b.

WASP-4b was discovered by \citet{wilson_etal2008}. It is a
1.12\,M$_{\rm J}$ hot Jupiter on a 1.34\,day orbit around a G7V star,
with a heavily irradiated atmosphere and large radius. Additional
orbital and physical parameters were measured also by
\citet{gillon_etal2009}, \citet{winn_etal2009}, and
\citet{southworth_etal2009}. Recently, \citet{beerer_etal2011}
performed photometry, using the IRAC instrument on the Warm Spitzer,
of the planet WASP-4b in the $3.6$ and $4.5$\,$\mu$m bands. Their data
suggests the WASP-4b atmosphere lacks a strong thermal inversion on
the day-side of the planet, an unexpected result for an highly
irradiated atmosphere.


WASP-4b was one of the pilot targets for our program because of its
short period and large size with respect to most transiting
exoplanets, with a relatively large predicted secondary eclipse
amplitude of about a millimagnitude, placing it well within our
expected range of detection. In Section 2 we discuss the observations
and their analysis. Section 3 presents the detection of the secondary
eclipse, and gives the measured brightness temperature for WASP4-b,
comparing it with theoretical models. Finally, Section 4 summarizes
the results and lists some conclusions.

\section{Observations and data reduction}

The $K_S$ band WASP-4b observations were obtained with the Infrared
Spectrometer And Array Camera \citep[ISAAC][]{moorwood_etal1998} at
the ESO VLT, on September 27, 2009. The observations were carried out
in the {\it FastPhot} mode, as in Paper~I\nocite{caceres_etal2009}. The InSb
1024$\times$1024\,px Aladdin detector was windowed down to
304$\times$608\,px, yielding a field of view of
$\sim$45$\times$90\,arcsec. The field of view includes WASP-4 and one
reference star -- R.A.\,23\,34\,18.4, Dec.\,-42\,04\,51.0 (J2000).
2MASS 
\citep{skrutskie_etal2006} lists for them respectively: $J$=11.2,
$H$=10.8, and $K_S$=10.7\,mag and $J$=11.7, $H$=11.3, and
$K_S$=11.2\,mag.

This observation mode generates a series of data-cubes with a fixed 
number of frames. Each frame is an individual detector integration, 
and there is nearly zero ``dead'' time between the integrations. The 
cubes are separated by an interval of $\sim$6\,sec for fits header 
merging, file transfer, and saving on a hard disk. For these 
observations the cubes had 250 frames, corresponding to integrations 
of 0.6\,sec each. In total, we collected 30,003 frames during a 
continuous 310\,min long run. 

\begin{figure}[!t]
  \begin{center}
    \includegraphics[scale=0.45]{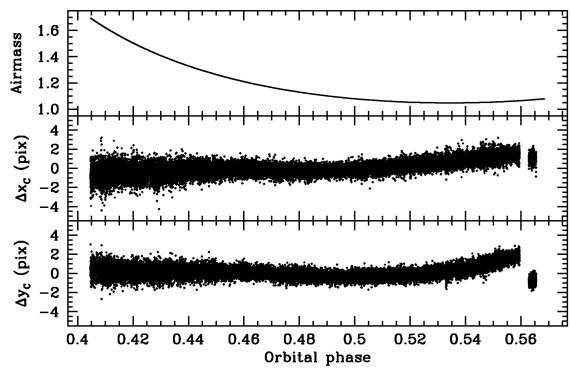}
  \end{center}
  \caption{\label{fig:center_amass} {\sl Upper panel:} The range of
    airmass covered during the run. {\sl Mid and bottom panels:} The
    variation around the median values of the position of the centroid
    of the stellar psf on the detector. The positions of those images
    discarded because of pointing jumps whose amplitude is larger than
    the plotted range are not shown, for the sake of clarity.}
\end{figure}

Standard procedures for NIR data reduction as dark subtraction and
flat fielding were performed on the data set. We applied aperture
photometry, using the IRAF\footnote{IRAF is distributed by the
  National Optical Astronomy Observatories, which are operated by the
  Association of Universities for Research in Astronomy, Inc., under
  cooperative agreement with the National Science Foundation.}
package {\it Daophot}, to measure the apparent fluxes of the two stars
in the field. An aperture radius of 1.92\,arcsec was selected
empirically to have the best compromise between the r.m.s. on the
transit light curve, determined as ratio of the target-vs-reference
fluxes, and the systematic noise introduced by a varying sky. The sky
background was estimated in a circular annulus with an inner radius of
3.10\,arcsec and outer radius of 7.5\,arcsec, from the centroid
of the point-spread function.

Occasionally, during the observations the central peak of the stars
exceeded the linearity limit of the detector of $\sim$6500\,ADU. To
avoid the non-linearity effects that may distort the occultation depth
measurement we discarded those frames. Moreover, at the end of the run
the detector was moved yielding in jumps in the position of the stars. All
the points after the first jump in position were also discarded. The
starting points showed also a jump that could not be corrected, and
these points were also removed to avoid including a systematic error
to the determined parameters. After cleaning the light curve, we leave
a total of 24,070 measurements, that were used during the subsequent
analysis.

The light curve shows a smooth trend during the period of observation,
most likely due to airmass variation, important during the first half 
of the night, and due to target drift across the detector, which 
dominated the end of the run, as shown in Fig. \ref{fig:center_amass}.
To remove these effects we modeled the base-line of the light curve with a polynomial 
that includes a linear dependency on time ($t$), airmass (sec\,$z$), 
and positions of the centroid of the star on the detector ($x_C$ and
$y_C$) as follows:
\begin{equation}\label{eq:polinorm}
f_{Bline} = a_0 + a_1 t + a_2 \sec\,z + a_3 x_C + a_4 y_C~.
\end{equation}

The coefficients in Eq.~\ref{eq:polinorm} were determined using the
entire light curve, with the fitting procedure described below. The
form of the polynomial was selected to flatten the out-of-eclipse
lightcurve after multiple experiments. We also attempted to include
other parameters, i.e. the full-width-half-maximum of stars on each
frame, but the correlation of the signal with this parameter was
negligible.

\section{Analysis and discussion}
Various parameters can be measured from the secondary eclipse light
curve. The thermal planetary emission is proportional to the eclipse
depth $d$, the orbital eccentricity $e$ and argument of periastron $w$
could be inferred through the analysis of both the eclipse central
time $T_C$, and the secondary eclipse length $\tau_{\rm occ}$. We
create an occultation model following the algorithm for uniform
illuminated transit light curve (i.e. neglecting the limb-darkening
contribution) from \citet{mandel_agol2002}, and we scaled the model
transit depth to fit the observed eclipse depth. The stellar and
planetary parameters for WASP-4 and WASP-4b were adopted from
\citet{winn_etal2009}: period $P$=1.33823214\,d, planet-to-star radius
ratio $p$=0.15375, semi-major axis $a$=5.473\,R$_\star$, orbital
inclination $i$=88.56\,deg, and stellar radius
$R$$_\star$=0.912\,$R$$_{\odot}$.

\begin{figure}[!t]
  \begin{center}
    \includegraphics[scale=0.45]{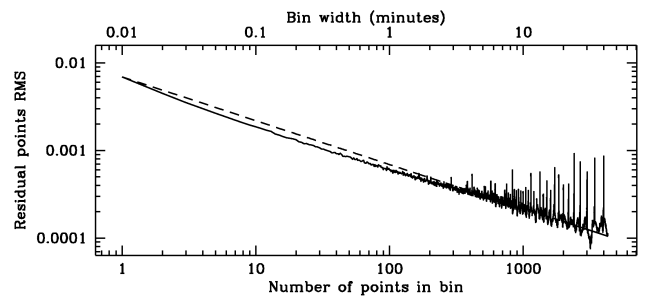}
  \end{center}
  \caption{\label{fig:rednoise} The effect of rednoise on the obtained
    light curve is shown. The total noise for a given bin (solid
    line), is compared with the expected Poisson noise (dashed
    line). The residual points for the complete light curve were used
    in this calculation.}
\end{figure}

\begin{figure}[!t]
  \begin{center}
    \includegraphics[scale=0.5]{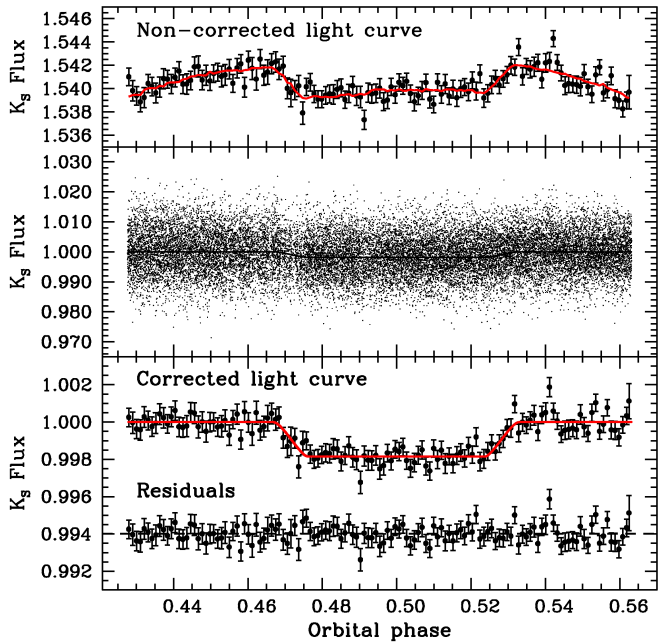}
  \end{center}
  \caption{\label{fig:lightcurve} The upper panel shows the 2-min
      bin uncorrected light curve, with the overlaid best fitting
      model considering the occultation model and the base-line
      model. The corrected $K_S$ light curve of WASP-4b is shown in
      the middle panel, including the complete data set, where the
      resulting rms in the out-of-eclipse points is $0.0069$. In the
      bottom panel each data point corresponds to a 2-minute bin,
      along with its error, and the best-fitting
      occultation model. The residuals for this curve are shown with
      an offset of $0.006$, and present a r.m.s. of $0.00072$. A color
      version of this figure is available in the electronic form.}
\end{figure}

Past experience with NIR data showed that correlated noise could
degrade photometric accuracy in these wavelengths. To evaluate the
contribution of this kind of noise on our photometry, we analyzed the
behavior of the light curve r.m.s. for different binning factors,
after removing the best-fitting model. It shows a small deviation from
the expected curve for pure Poisson noise (Fig. \ref{fig:rednoise})
over range of binning corresponding to times of the order of tens of
minutes -- similar to the duration of the ingress-egress for the
secondary eclipse.

The best fitting model parameters were found with the multidimensional
minimization algorithm AMOEBA \citep{press_1992}, where the free
parameters were the eclipse depth, central time, length, and the
coefficients of the base-line in Eq. \ref{eq:polinorm}, on the
  2-min binned light curve, yielding 131 data points. The function to
minimize was the $\chi^2$ statistic, for a model that is:

\begin{equation}\label{eq:model}
mod = m_{occ} \times f_{Bline} ~,
\end{equation}

where $m_{occ}$ corresponds to the occultation model described
above. To determine the errors in the fitted parameters, we performed
the bootstrapping procedure described in Paper
I\nocite{caceres_etal2009}: we first subtracted the best fitting model
from the data set, then we took the set of residuals and shifted the
$i^{th}$ residual to become the $i+1$$^{th}$ residual, and the last to
become the first. Next, we added the re-ordered residuals to the best
fitting model and run it through the $\chi^2$ minimization procedure
instead of the data to obtain a new set of fitting results. This
procedure was repeated until a full circle over the ``good'' points
was completed, and the reported uncertainties were defined by the
68.3\% level around the median values of the distributions. Therefore,
our errors properly account for the correlated noise in the data set.

\begin{figure}[!pt]
  \begin{center}
    \includegraphics[scale=0.45]{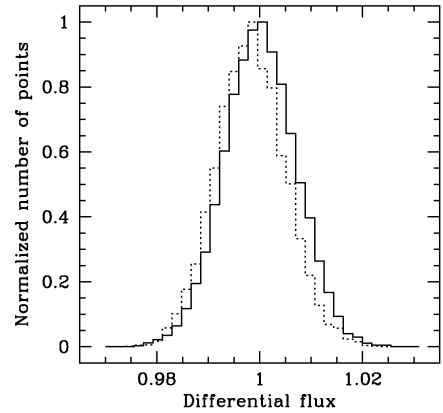}
  \end{center}
  \caption{\label{fig:depth} Distribution of the flux during the
    eclipse (dashed line) and out of the eclipse (solid line). The bin
    width is $0.185\%$, equal to the calculated eclipse depth.}
\end{figure}

The final corrected light curve is shown in
  Fig.~\ref{fig:lightcurve}.  Table~\ref{table:data} lists the
  individual flux measurements, with their time-stamps and Poisson
  uncertainties. The resulting parameters and their uncertainties for
  WASP-4b are listed in Table~\ref{table:params}.

\begin{table}[t]
\caption{Relative photometry of WASP-4b. Extra digits are 
given to avoid round up errors.}
\label{table:data}
\begin{tabular}{ccc}
\hline
\hline
HJD & Relative flux & Uncertainty\\
\hline
2455102.5150000 & 0.99951 & 0.00427\\
2455102.5150069 & 0.99732 & 0.00427\\
2455102.5150138 & 1.00339 & 0.00421\\
2455102.5150277 & 0.99452 & 0.00424\\
2455102.5150347 & 1.00043 & 0.00426\\
\hline
\end{tabular}\\[5pt]
\begin{minipage}[t]{0.8\columnwidth}
Only a small portion of the data set is presented in this table, to
exemplify its presentation format. The complete set can be found in
the electronic version of the Journal.
\end{minipage}
\end{table}

The significance of our detection can be assessed from
Fig.~\ref{fig:depth}, where the distribution of the relative flux
(with respect to the reference star) during the eclipse and outside of
the eclipse are compared. The bin width is equal to the calculated
eclipse depth $d = 0.185$\%, and each distribution curve was
normalized for comparison purposes. A K-S test applied on these
distributions shows they are identical with a $\sim99\%$ of
probability, but displaced exactly one bin.

\begin{table}[ht]
\caption{Derived parameters of WASP-4b.}
\label{table:params}
\begin{tabular}{cccc}
  \hline
  \hline
  Parameter & Value & 68.3\% Confidence Limits & Unit\\
  \hline\\[0.0pt]
  $T_C$         & 2455102.61162 & -0.00071, +0.00077  & HJD \\
  $d$           & 0.185    & -0.014, +0.013  & \% \\
  $\tau_{\rm occ}$ & 0.0881     & -0.0017, +0.0022   & d \\
  $e\cos\omega$ & 0.0027       & $\pm$ 0.0018     &  - \\
  $T_B$         & 1995         & $\pm$ 40          & K \\
  \vspace{3pt}\\
  \multicolumn{4}{c}{Base-line parameters}\\
  $a_0$         & 1.6933 & -0.0169, +0.0214 &\\
  $a_1$         & -0.0364 & -0.0027, +0.0042 &\\
  $a_2$         & -0.0112 & -0.0014, +0.0013 &\\
  $a_3$         & -0.00025 & -0.00074, +0.00062 &\\
  $a_4$         & -0.00175 & -0.00060, +0.00078 &\\
\hline
\end{tabular}
\end{table}

The center of the secondary eclipse occurs at $\phi = 0.49933
  \pm 0.00059$, assuming the ephemeris of
\citet{southworth_etal2009}. For zero eccentricity
\citep{wilson_etal2008, gillon_etal2009, winn_etal2009,
  southworth_etal2009}, and a light travel time of $23.2$\,sec for
this system \citep{loeb_2005}, the secondary eclipse is expected to
occur at phase $\phi_{exp}=0.5002$. The phase difference $\delta \phi$
implies a non-zero eccentricity, that could be determined knowing the
primary transit length $\tau_{\rm tr}$, from the equations
\citep{charbonneau_etal2005}:

\begin{equation}
  e\cos\omega \simeq \pi~\delta\phi ~ ,
\end{equation}
\begin{equation}\label{eqn:ell}
  e\sin\omega \simeq \frac{\tau_{\rm tr} - \tau_{\rm occ}}{\tau_{\rm tr} + \tau_{\rm occ}} ~.
\end{equation}

The accuracy of our eclipse length measurement prevents us from using
Eq. \ref{eqn:ell}, so we only use the former equation, to put a upper
limit on the eccentricity $e\cos\omega = 0.0027 \pm 0.0018$. Within
the 3-$\sigma$ level, this results argues in favor of a nearly
circular orbit.

\begin{figure*}[t]
  \begin{center}
    \includegraphics[scale=0.8]{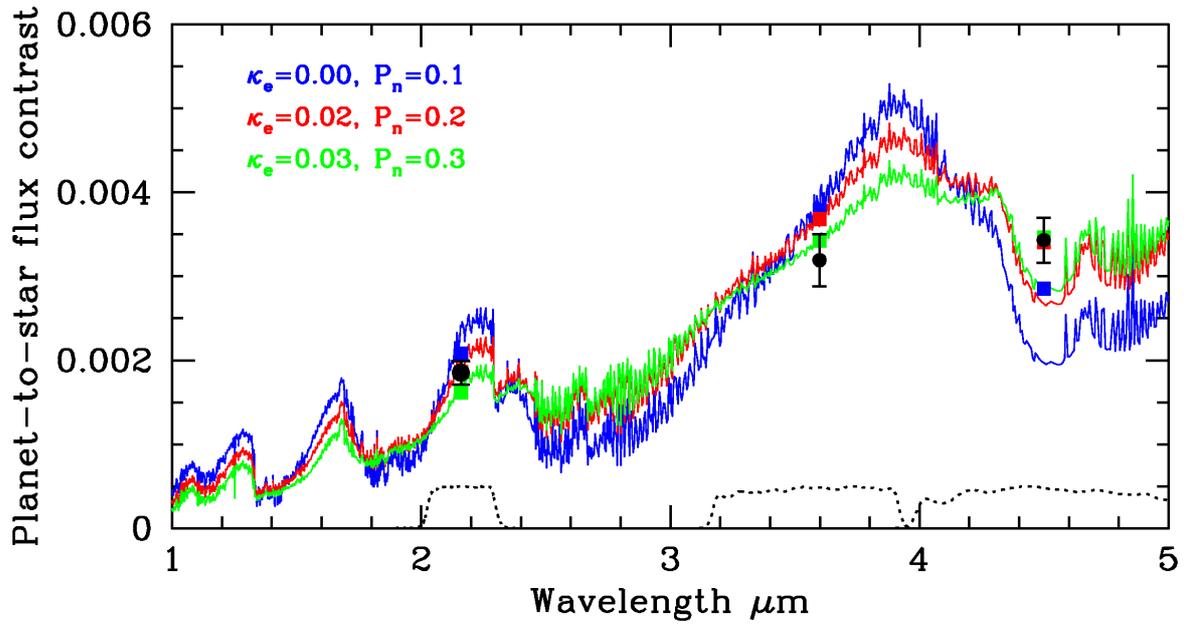}
  \end{center}
  \caption{\label{fig:model} The planet-to-star flux contrast as a
    function of wavelength. The solid lines show three models with
    $\kappa_e=0.00-0.03$\,cm$^2$/g, and $P_n=0.1-0.3$ for the planet
    WASP-4b. The squares are the estimated band-weighted measurements
    for these models, and the black point at $2.16\,\mu$m represents
    our measurement, with its error bars. Note that the red model
      point at $2.16\,\mu$m is overlapped by our measurement. The
    measurements in \citet{beerer_etal2011} are also shown at $3.6$,
    and $4.5\,\mu$m. The transmission curves for the three filters are
    also shown for comparison as dotted lines. A color version of
      this figure is available in the electronic form.}
\end{figure*}

Previous planet-to-star contrast measurements for WASP-4b have been
obtained by \citet{beerer_etal2011} in the $3.6$ and $4.5\,\mu$m broad
band filters with the Spitzer Space Telescope. As a zero-order
approximation, we represent the planet with a black body with
temperature $T_B$, and the host star with a
\citet{hauschildt_etal1999} stellar atmosphere model, with parameters
$T_{eff} = 5500\,K$, $\log g = 4.5$, and [Fe/H]\,=\,0.0
\citep{gillon_etal2009}. They are divided to obtain the planet-to-star
flux contrast as a function of wavelength.  The expected depth was
calculated as weighted average of the contrast curve, weighted by the
$K_S$ filter transmission curve, the atmospheric transparency, and the
detector efficiency curve. The best-fitting black body curve yielded a
planetary brightness temperature of $T_B = 1995 \pm 40$\,K at
$\sim$2.2\,micron.

The observational data of highly irradiated planets indicate very low
Bond albedo $A_B$ values \citep[e.g.][]{charbonneau_etal1999,
  rowe_etal2008, rogers_etal2009}, implying that the atmospheres of
those planets are highly heated by the stellar radiation, and probably
bloated. The temperature at which the planet re-emits the energy
absorbed from the stellar flux, at a given distance $a$ from the star,
in units of the stellar radius, is given by $T_{\mathrm{eq}} = T_{\rm
  eff}\,a^{1/2} \left(f (1 - A_B\right))^{1/4}$. Here, the $f$ factor
refers to the fraction of energy that is re-radiated from the planet
to the observer. If the stellar radiation absorbed by the planet is
re-radiated isotropically then $f = 1/4$, and we could determine for a
zero Bond albedo an equilibrium temperature for the planet of
$T_{\mathrm{eq}} = 1662$\,K.  The difference between the equilibrium
and the brightness temperatures suggests a poor energy redistribution
in the atmosphere of WASP-4b. In fact, to obtain the calculated
brightness temperature the re-radiation factor should be $f=0.52$,
indicating a planet with re-radiation only from the day-side
\citep{burrows_etal2008}, when assuming a zero Bond-albedo.  This
result is in agreement with the results in \citet{beerer_etal2011},
who found that the \citet{fortney_etal2008} models with a small
redistribution of heat, and an absence of TiO in the WASP-4b
stratosphere, are best fits for their measurements.

Planetary atmosphere models tuned for WASP-4b are shown in
Fig. \ref{fig:model}. The modeled atmospheres assume radiative and
chemical equilibrium, and employ the chemical compositions and
thermo-chemistry found in \citet{burrows_sharp1999}, and
\citet{sharp_burrows2007}, and the opacities described in
\citet{sharp_burrows2007}, and references therein. These models were
calculated for different redistribution factors $P_n$, where $P_n = 0$
implies no redistribution, and $P_n = 0.5$ means maximum
redistribution \citep{burrows_etal2008}. This parameter represents the
fraction of energy that is transferred from the day-side to the
night-side of the planet. Moreover, some highly irradiated atmospheres
require the inclusion of an extra absorber to fit calculated eclipse
depths, this accounting for the presence of inversion layers in the
upper atmosphere \citep{burrows_etal2007, knutson_etal2009}. This
scenario was tested by introducing an opacity parameter $\kappa_e$ for
this extra absorber as a second parameter to explore. 

Our observations favor lower $P_N$ parameter values, arguing in favor
of an inefficient redistribution of the received radiation from the
day-side to the night-side of the planet, as shown in
Fig. \ref{fig:model}. The models best suited with our measurement
correspond to atmospheres with null or low amounts of a stratospheric
absorber, this leading to non inverted, or scarcely inverted
atmospheres. A similar result was found in \citet{beerer_etal2011},
for observations in the mid-IR. When considering the three measurements,
i.e. the ISAAC one at 2.2\,$\mu$m, and the Spitzer ones at $3.6$ and
$4.5\,\mu$m, the best fitting models is the one with an inefficient
redistribution of heat, and an small quantity of absorber in the
planet stratosphere. Thus, the lack of a strong thermal inversion in
the atmosphere of WASP-4b could be inferred from this IR detection.

\section{Conclusions}
The ultrafast photometry technique developed in Paper
I\nocite{caceres_etal2009} is able to detect very low amplitude
secondary eclipses of extrasolar planets. We have measured the
secondary eclipse depth of WASP-4b which has an amplitude of $d_{K_S}
= 0.185$\% at $\geq$10$\sigma$ level. This secure detection
strengthen our confidence in the new method, and indicates that it
could be fruitful to attempt the observation of secondary eclipses at
shorter wavelengths, where the eclipses are shallower, but the NIR
signal is still within our observational capabilities.

For the stellar and planetary parameters, this yields a brightness
temperature of $T=1995$\,K in the $K_S$, which agrees with the
specific models for WASP-4b, and argues in favor of inefficient heat
redistribution from the day-side to the night-side of the planet. The
absence of an strong thermal inversion in the stratosphere of WASP-4b
is inferred from our near-IR measurement, and mid-IR measurements in
the literature.

\begin{acknowledgements}
  This work is supported by ESO, by BASAL Center for Astrophysics and
  Associated Technologies PFB-06, by FONDAP Center for Astrophysics
  15010003, and by Ministry for the Economy, Development, and
  Tourism's Programa Inicativa Cient\'{i}fica Milenio through grant
  P07-021-F, awarded to The Milky Way Millennium Nucleus. A.B would
  like to acknowledge support in part by NASA grant NNX07AG80G and
  through JPL/Spitzer Agreements 1328092, 1348668, and 1312647. The
  authors would like to acknowledge David Anderson by his useful
  comments and suggestions.
\end{acknowledgements}

\bibliographystyle{aa}
\bibliography{paper2}

\end{document}